\documentstyle[preprint,epsfig,aps]{revtex}

\epsfverbosetrue

\newcommand{\beq}{\begin{equation}}
\newcommand{\enq}{\end{equation}}

\newcommand{\Ar}[1]{\mbox{Ar$_{#1}$}}
\newcommand{\Si}[1]{\mbox{Si$_{#1}$}}

\newcommand{\cpl}{Chem. Phys. Lett.}
\draft

\begin{document}
\title{Structures of Silicon Clusters}
\author{Jun Pan,\footnote{Also at the Department of Physics,
New York University, New York, New York 10003-6621}
Atul Bahel,
and Mushti V. Ramakrishna}
\address{The Department of Chemistry, New York University,
New York, New York 10003-6621.}
\date{Surface Review and Letters, in press \today}
\maketitle

\begin{abstract}

We determined the structures of silicon clusters in the 11-14 atom size
range using the tight-binding molecular dynamics method.   These
calculations reveal that \Si{11} is an icosahedron with one missing
cap, \Si{12} is a complete icosahedron, \Si{13} is a surface capped
icosahedron, and \Si{14} is a 4-4-4 layer structure with two caps.  The
characteristic feature of these clusters is that they are all surface.

\end{abstract}
\vspace{0.1 in}
\pacs{PACS numbers: 36.40.+d, 61.43.Bn, 61.46.+w, 68.35.Bs, 82.65.My}

The size dependence of the structural, electronic, optical, and
chemical properties of clusters has been the subject of experimental
and theoretical studies in both physics and chemistry communities for
more than a decade \cite{Sugano:91}.    Recently, the covalent
semiconductor clusters have received substantial attention because of
their expected importance in microelectronics industry
\cite{Elkind:87,Jarrold:89,Phillips:88,Jelski:88,Kaxiras:89,Bolding:90,Patterson:90,Swift:91,Roth:94,RK:91,RK:94}.
Since silicon is technologically the most important semiconductor
material, these studies have attempted to understand the evolution of
its structural and optical properties as a function of cluster size.

Because silicon clusters are highly reactive, they are mostly
synthesized in a molecular beam under high vacuum conditions
\cite{Elkind:87,Jarrold:89}.  Consequently, the structures of these
clusters are unknown experimentally.  Theoretical calculations have now
established the structures of \Si{N} clusters in the $N$ = 2-10 atom
size range \cite{Krishnan:85}.  In order to extend this work, we have
recently carried out tight-binding molecular dynamic (TB-MD)
simulations on \Si{N} clusters in the $N$ = 11-14 atom size range.  The
results of these simulations are reported here.

The choice of the tight-binding method for the study of the
cluster structures is motivated by its accuracy and computational
efficiency.  Toma\'nek and Schl\"uter have shown that a properly
constructed tight-binding Hamiltonian can yield cluster structures as
accurately as the first principles methods \cite{Tomanek:86}.  More
recently, Menon and Subbaswamy have constructed a highly accurate
tight-binding Hamiltonian for silicon clusters
\cite{Menon:93-2,Ordejon:94}.  This Hamiltonian
includes Harrison's universal parameters appropriate for the
description of bulk Si \cite{Harrison:80}, supplemented with two to
four additional parameters for the description of the silicon
clusters.  These additional parameters are derived by fitting to the
\Si{2} bond length and vibrational frequency and to the overall
size-dependent cohesive energy curve of clusters
\cite{Menon:93-2,Ordejon:94}.  Most importantly, none of the parameters
of this Hamiltonian are fit to any of the cluster structures.  Full
details of the Hamiltonian are described elsewhere
\cite{Menon:93-2,Ordejon:94}.

As the size of the cluster grows, the number of structural isomers
increases exponentially, with the result that searching the complete
configuration space for the global potential energy minimum becomes a
formidable task.  We found that by combining the tight-binding method
with the molecular dynamics simulated annealing technique we can
efficiently search the cluster configuraton space and determine the
ground state geometry.  We used this TB-MD method in all our
calculations reported here.

Figures 1-4 display the lowest energy structures obtained using this
TB-MD technique for clusters in the $N$ = 11-14 atom size range.  The
\Si{11} structure displayed in Fig. 1  consists of two tetragons in the
anti-prism geometry and three caps.  Two of these caps are attached to
the opposite faces of the two tetragons,
while the third one is attached to the edge
of the top tetragon.  Rohlfing and Raghavachari found a similar
structure to be a possible candidate structure for the ground state of
\Si{11} \cite{Rohlfing:90}.  The edge cap in our structure is replaced
by a second face cap in the structure of Rohlfing and Raghavachari
\cite{Rohlfing:90}.  We investigated both these structures and found
that our structure is more stable by 0.27 eV than the
Rohlfing and Raghavachari structure.  Finally, if we
remove the edge cap from the top and relax the resulting structure we
obtain one of the low energy \Si{10} structures.  Hence, we may view
\Si{11} as a continuation of the \Si{10} structure.  Similarly, if we
add an edge cap to the bottom tetragon we would obtain the twelve
atom icosahedral cage structure.  Consequently, we may also view the
\Si{11} structure as an incomplete icosahedral cage structure.

In Fig. 2 we display the structure of \Si{12} structure.  This
structure is an icosahedral cage.  Such a spherical cage structure has
not been predicted or observed for any of the 12-atom elemental
clusters.  Adding
a face cap to \Si{12} gives the lowest energy \Si{13} structure, as
shown in Fig. 3.  An alternative structure, derived by placing a Si
atom inside the cage of \Si{12}, is a high energy local minimum.
Similarly, adding a cap to the \Si{13} structure does not yield the
lowest energy \Si{14} structure.  As displayed in Fig. 4, \Si{14}
assumes a layer structure consisting of three planes of four atoms each
and two adjacent face caps.  By suitably rotating this structure, we
may also describe it as a pentagon sandwich (or prism) with two caps
each at the top and the bottom.  The pentagon prism is somewhat
distorted and displaced.  We also considered a bi-capped hexagonal
anti-prism as a candidate for the ground state of \Si{14}.  However,
this structure proved to be unstable, indicating that six-atom ring
structures are still not favored in these small clusters.

The Ar$_N$ clusters in this size regime assume different structural
pattern than the corresponding \Si{N} clusters \cite{Liu:94}.  For
example, Ar$_{13}$ is an icosahedron but with an atom inside the cage
\cite{Liu:94}.  Furthermore, if the central Ar atom is removed the
resulting Ar$_{12}$ structure is unstable.  The stable Ar$_{12}$ and
Ar$_{11}$ cluster structures may be derived from Ar$_{13}$ by removing
one and two surface atoms, respectively \cite{Liu:94}.  Similarly,
Ar$_{14}$ and Ar$_{15}$ structures may be derived from Ar$_{13}$ by
adding one and two face caps, respectively, to its surface
\cite{Liu:94}.  Hence, we may view the structures of Ar$_N$ ($N$ =
9-15) clusters in the neighborhood of \Ar{13} as being derived from
\Ar{13} by the addition or subtraction of surface atoms.  Such a simple
structural relationship exists for \Si{N} clusters only in a very
limited range of cluster sizes.  For example, \Si{11} and \Si{13}
cluster structures may be derived from the \Si{12} structure by the
removal or addition of one surface atom.  However, \Si{10} and \Si{14}
structures are not so simply related to the structure of \Si{12}.
Similarly, surface caps stabilize \Si{N} structures much more than the
interior caps.  In sharp contrast, removing the interior atom from the
\Ar{N} ($N$ = 9-15) clusters make the resultant structures unstable.
Hence, the \Si{N} structures are quite different from those of the
corresponding Ar$_N$ clusters.  Since, two-body interaction is the
dominant part of the Ar-Ar potential, the differences between Ar$_N$
and \Si{N} cluster structures are due to the many-body interactions of
the Si-Si potential and the directional nature of the covalent
bonding.  The directional bonds in \Si{N} stabilize the cage structure,
while the isotropic Ar-Ar pair potentials stabilize closed shell
structures with maximum coordination.

Previous work on silicon clusters in this size regime employed both
classical potentials \cite{Barojas:86,Feuston:87,Chelikowsky:89} as
well as first principles methods
\cite{Krishnan:85,Rohlfing:90,Roth:92}.  Our calculated cluster
structures in the $N$ = 2-10 atom size range agree with the first
principles calculations of Raghavachari \cite{Krishnan:85}.
Furthermore, our structure for \Si{11} is similar to the one obtained
by Rohlfing and Raghavachari \cite{Rohlfing:90}.  However, our
structure for \Si{13} is  qualitatively different from that obtained by
R\"othlisberger and Andreoni \cite{Roth:92}.  The structures of \Si{12}
and \Si{14} have not been investigated using the first principles
methods.

Classical calculations exist on the structures of all these
clusters.  The calculations of Barojas and Levesque \cite{Barojas:86}
and Feuston and co-workers \cite{Feuston:87} employed the
Stillinger-Weber potential \cite{Stillinger:85}.   However, this
potential almost always yielded incorrect cluster structures because
it  was derived by fitting to bulk Si properties.  Indeed, the
predicted structures of \Si{11} and \Si{13} using this potential
disagree with our results as well as with those of the first-principles
methods.  Chelikowsky also reported the structures of \Si{13} and
\Si{14} structures using his own classical potential that was designed
specifically for clusters \cite{Chelikowsky:89}.  However, his
prediction of atom-centered icosahedron for \Si{13} is in disagreement
with our calculations as well as with that of R\"othlisberger and
Andreoni.  Likewise, Chelikowsky's structure for \Si{14} is quite
different from our structure reported here.  There are no first
principles calculations on \Si{14}.  This comparison shows that our
tight-binding method gives results consistently different from those of
classical potentials, while agreeing quite well with the calculations
of first-principles methods.

There is a concern that the tight-binding Hamiltonian we employ may be
behaving more like an isotropic pairwise additive potential.  If that
were the case, we would have obtained atom centered icosahedral
structure for \Si{13}, as is the case for Ar$_{13}$ \cite{Liu:94}.
However, it turns out that the atom centered icosahedron is a high
energy local minimum for \Si{13}.  Furthermore, our calculated
structures for clusters in the $N$ = 2-10 atom size range are in good
agreement with the first principles calculations \cite{Krishnan:85}.
These two observations confirm that our tight-binding Hamiltonian
correctly incorporates the many-body interactions necessary for the
accurate description of the cluster structures.

In summary, our tight-binding molecular dynamic simulations predict
unforeseen structures in the $N$ = 11-14 atom size range.  The \Si{11}
cluster is an incomplete icosahedron, \Si{12} is an icosahedral cage,
\Si{13} is a surface capped icosahedron, and \Si{14} is a bi-capped
4-4-4 layer structure.  Since most of these are low symmetry
structures, they may be described differently by viewing them through
different orientations.  All these clusters have low energy isomers
that may play an important role in the observable properties.  While
\Si{11-13} are close to being spherical, \Si{14} is a prolate
ellipsoid.  The \Si{12} cluster in particular assumes highly spherical
icosahedral cage structure.  Such a spherical cage structure has not
been predicted or observed for any of the silicon clusters.  The most
characteristic feature of all these clusters is that they are all
surface.

This research is supported by the New York University and the Donors of
The Petroleum Research Fund (ACS-PRF Grant \# 26488-G), administered by the
American Chemical Society.

\begin{figure}
\caption{The lowest energy structure of \Si{11} cluster.  It is an
incomplete icosahedron.}

\end{figure}

\begin{figure}
\caption{The lowest energy structure of \Si{12} cluster.  It is an
icosahedron.
}

\end{figure}

\begin{figure}
\caption{The lowest energy structure of \Si{13} cluster.  It is a
surface capped icosahedron.
}

\end{figure}

\begin{figure}
\caption{The lowest energy structure of \Si{14} cluster.  It is a
layer structure consisting of three planes of four atoms each
and two adjacent face caps.
}

\end{figure}

\end{document}